\documentclass[12pt,draftclsnofoot,onecolumn]{IEEEtran}
\IEEEoverridecommandlockouts
\usepackage[utf8]{inputenc}
\usepackage{bm}
\usepackage{amsfonts}
\usepackage{amssymb}
\usepackage{amsmath}
\usepackage{algorithm}
\usepackage{algpseudocode}
\usepackage[pdftex]{graphicx}
\usepackage{siunitx}
\usepackage{cite}
\usepackage[font=small]{caption}
\usepackage{multicol}

\title{Rate Optimization for RIS-aided mMTC Networks in the Finite Blocklength Regime}
\author{Sergi~Liesegang,~\IEEEmembership{Student Member,~IEEE}, Alessio~Zappone,~\IEEEmembership{Senior Member,~IEEE}, Olga~Mu\~noz,~\IEEEmembership{Member,~IEEE}, and Antonio~Pascual-Iserte,~\IEEEmembership{Senior Member,~IEEE}
\thanks{This work was supported by the project ROUTE56 (PID2019-104945GB-I00 / MCIN / AEI / 10.13039/501100011033) funded by the Agencia Estatal de Investigaci\'on (Spanish Ministry of Science and Innovation); and the FPI grant BES-2017-079994, funded by the Spanish Ministry of Science, Innovation, and Universities.}
\thanks{Alessio Zappone is with the Department of Electrical and Information
Engineering, University of Cassino and Southern Lazio, 03043 Cassino, Italy
(e-mail: alessio.zappone@unicas.it)}
\thanks{Sergi Liesegang, Olga Mu\~noz, and Antonio Pascual-Iserte are with the Department of Signal Theory and Communications,
Universitat Polit\`ecnica de Catalunya, 08034 Barcelona, Spain (e-mails: sergi.liesegang@upc.edu, olga.munoz@upc.edu, antonio.pascual@upc.edu).

\copyright 2023 IEEE. Personal use of this material is permitted. Permission from IEEE must be obtained for all other uses, in any current or future media, including reprinting/republishing this material for advertising or promotional purposes, creating new collective works, for resale or redistribution to servers or lists, or reuse of any copyrighted component of this work in other works.

DOI: 10.1109/LCOMM.2022.3231717

}}

\begin{document}

\markboth{Accepted Paper at IEEE Communications Letters (vol. 27, no. 3, Mar. 2023)}
{}

\maketitle

\begin{abstract}
Reconfigurable intelligent surfaces (RISs) have become a promising candidate for the development of future mobile systems. In the context of massive machine-type communications (mMTC), a RIS can be used to support the transmission from a group of sensors to a collector node. Due to the short data packets, we focus on the design of the RIS for maximizing the weighted sum and minimum rates in the finite blocklength regime. Under the assumption of non-orthogonal multiple access, successive interference cancelation is considered as a decoding scheme to mitigate interference. Accordingly, we formulate the optimizations as non-convex problems and propose two sub-optimal solutions based on gradient ascent (GA) and sequential optimization (SO) with semi-definite relaxation (SDR). In the GA, we distinguish between Euclidean and Riemannian gradients. For the SO, we derive a concave lower bound for the throughput and maximize it sequentially applying SDR. Numerical results show that the SO can outperform the GA and that strategies relying on the optimization of the classical Shannon capacity might be inadequate for mMTC networks.
\end{abstract}

\begin{IEEEkeywords}
Massive machine-type communications, reconfigurable intelligent surfaces, finite blocklength regime, gradient ascent, sequential optimization, semi-definite relaxation.
\end{IEEEkeywords}

\section{Introduction} \label{sec:1}
Massive machine-type communications (mMTC) play an essential role in the next generation of cellular networks \cite{Boc18}. They represent a type of setup where large sets of devices send their information to a base station or collector node (CN) in an unsupervised manner. Weather forecasting, surveillance systems, and health monitoring are only some examples of possible mMTC applications.

Given that mMTC transmissions consist of packets with short lengths, reliable communication can be sometimes difficult (especially in scenarios with poor channel propagation, e.g., millimeter-wave bands). That is why in this work, to boost the system performance, we consider the use of a reconfigurable intelligent surface (RIS) \cite{Hua19}. The RIS will be designed to maximize the throughputs of the mMTC terminals in the finite blocklength regime (FBLR) \cite{Pol10}.

A RIS can be described as a large passive surface that allows the adaptation to the wireless environment. In essence, these types of structures act as reflectors that can point the signals toward the target destination and enhance the received signal strength. This gain in received power, together with its low cost and easy deployment, make RISs one of the potential technologies for future mobile networks \cite{DiR20}.

Due to the vast connectivity and lack of resources in (delay-insensitive) mMTC, we adopt a non-orthogonal multiple access (NOMA) transmission and a decoding scheme based on successive interference cancelation (SIC) \cite{Dai18}. The role of the RIS will be to adapt the channel to the SIC procedure, which can help to reduce the influence of the interference along with the mitigation of channel quality drawbacks \cite{Jia20}. 

The design of the RIS for the data rate maximization results in non-convex problems. To circumvent that, the optimizations are addressed using: (i) gradient ascent (GA) and (ii) sequential optimization (SO) with semi-definite relaxation (SDR). Both techniques will allow us to find sub-optimal and feasible solutions, whose performance is evaluated numerically and whose computational complexity is also analyzed.

Regarding other works in the literature, some advances have been reported. The authors of \cite{Cao21} proposed a block coordinate descent method to maximize the sum rate in device-to-device communications. In \cite{Mu20}, the authors investigated the use of SIC for RIS-aided networks and maximized the total throughput with the help of convex relaxations and approximations. Similarly, the ergodic rate in the presence of correlated Rician fading was maximized in \cite{Xu21} using alternating optimization (AO). Nevertheless, to the best of the authors' knowledge, no studies have been conducted in the direction of RIS design for rate maximization in mMTC systems under the FBLR.

The remainder of this paper is structured as follows. Section~\ref{sec:2} describes the system model and the optimization problems. Section~\ref{sec:3} presents the proposed solutions. Section~\ref{sec:4} provides the simulation results. Section~\ref{sec:5} concludes the work. 

\section{System Model and Problem Formulation} \label{sec:2}
Throughout this paper, we will consider a setup with a set of $M$ single-antenna sensors connected to a  single-antenna CN. Each device maps its measurements into transmit symbols $x_i \sim \mathcal{CN}(0,P_i)$ and sends them to the CN on a NOMA basis.

\begin{figure}[t]
\centering
\includegraphics[scale = 0.5]{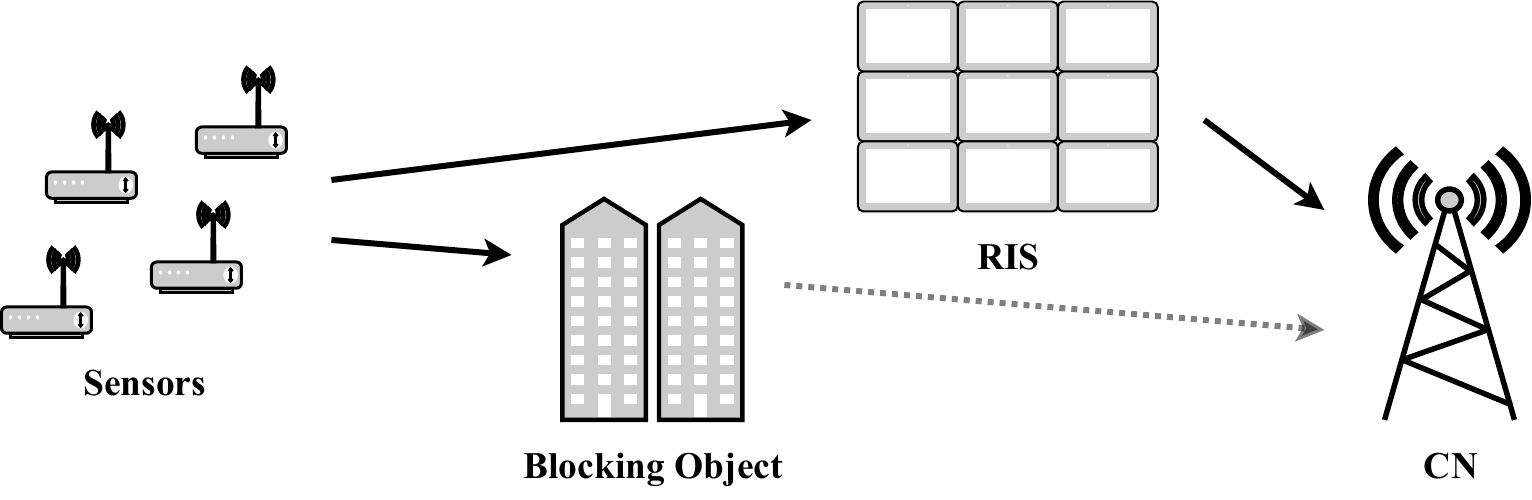}
\caption{Scenario with $M = 4$ sensors and $L = 9$ RIS elements. Solid and dotted lines indicate strong and weak paths, respectively.}
\label{fig:1}
\end{figure}

In order to support the transmission from the sensors to the CN, we incorporate the usage of a RIS with $L$ reflecting elements. By means of phase shifters, the RIS will be responsible for spatially focusing the different signals as needed \cite{Bjo20}. An illustrative example of a scenario with $M = 4$ and $L = 9$ is depicted in Fig.~\ref{fig:1}, where the direct path is blocked and the RIS is used to create an additional path to reach the CN.

The received signal at the CN can be expressed as
\begin{equation}
    y \triangleq \sum_{i = 1}^M \left(q_i + \bm{g}_{\textrm{R}}^{\textrm{T}}\bm{\Psi}\bm{g}_i\right) x_i + w,
    \label{eq:1}
\end{equation}
where $q_i \in \mathbb{C}$ is the channel from sensor $i$ to the CN, $\bm{g}_{\textrm{R}} \in \mathbb{C}^L$ is the channel between RIS and CN\footnote{For the purpose of exposition, this work focuses on a single-antenna CN to present the philosophy behind the RIS optimization without complicating unnecessarily the notation. Note, however, that the techniques proposed in this paper could be extended to other detection schemes based on SIC and a multiple-antenna CN, such as in \cite{Dai18}, where a set of beamformers are applied at the receiver, thanks to which the vector channels can be scalarized.}, $\bm{g}_i \in \mathbb{C}^L$ is the channel from sensor $i$ to the RIS, $\bm{\Psi} \triangleq \textrm{diag}\left(\lambda_1 e^{j\phi_1},\ldots,\lambda_L e^{j\phi_L}\right)$ is the RIS reflection matrix with $\lambda_l \in [0,1]$ denoting the amplitude coefficients and $\phi_l \in [0,2\pi)$ representing the phase-shifts \cite{Li21}, and $w \sim \mathcal{CN}(0,\sigma_w^2)$ is the thermal noise. 

As mentioned, in latency-tolerant mMTC services, SIC can be safely employed to alleviate the interference effects. In a nutshell, decoding is applied sequentially following a certain order, and, at each step, the previous signals are canceled, i.e., their contributions are subtracted from the received signal $y$. For simplicity, we assume the direct channel $q_i$ is negligible in comparison with the sensors-RIS-CN link \cite{Hua19,DiR20}. Hence, considering perfect cancelation and a fixed\footnote{ Note that the SIC ordering could also be optimized \cite{Dai18,Din20}, yet it involves a combinatorial problem whose solution is challenging and unclear in the FBLR. That is why its selection is left for future studies.} decoding order $1,\ldots,M$, the signal-to-interference-plus-noise ratio (SINR) when detecting the signal from sensor $i$ yields
\begin{equation}
    \rho_i \triangleq \frac{P_i \left\vert \bm{\psi}^{\textrm{T}} \bm{h}_i \right\vert^2 }{\displaystyle \sigma_w^2 + \sum_{j = i + 1}^M P_j \left\vert \bm{\psi}^{\textrm{T}} \bm{h}_j \right\vert^2},
    \label{eq:2}
\end{equation}
where $\bm{h}_i \triangleq \textrm{diag}(\bm{g}_{\textrm{R}}) \bm{g}_i$ is the (known\footnote{In practice, imperfect channel knowledge should be considered \cite{Bjo20}. Thus, as shown in Section~\ref{sec:4}, our results will serve as a performance upper bound.}) cascaded channel and $\bm{\psi} = [\psi_1,\ldots,\psi_L]^{\textrm{T}}$ is the vector containing the main diagonal elements of the RIS matrix $\psi_l \triangleq \lambda_l e^{j \phi_l}$ such that $\vert \psi_l \vert \leq 1$.

For the proper performance of the SIC procedure, it is imperative to have unbalanced received powers for separating the signals of the devices, i.e., power-domain NOMA \cite{Dai18}. In conventional wireless networks, this difference can be attained by modifying the sensors' transmit powers $P_i$, or by designing spatial filters at the CN (when equipped with multi-antenna technology). However, in mMTC scenarios, power allocation mechanisms are not feasible due to the large number of devices involved. Besides, the simplicity of the sensors often limits the number of power levels and their adjustment. That is why we will consider fixed $P_i$ and no transmit power control here. 

Given the presence of the RIS, the received powers can still be adjusted. Consequently, the role of the RIS is two-fold: (1) improve the signals' quality and strength, but also (2) adapt the channels to the SIC procedure (ensure unbalanced powers). Both factors will contribute to increasing the data rate.

In line with that, under the assumption of short data packets, the throughput of sensor $i$ (in bits/Hz) is given by \cite{Pol10}
\begin{equation}
    R(\rho_i,n_i,\epsilon_i) \triangleq C(\rho_i) - \sqrt{\frac{V(\rho_i)}{n_i}} Q^{-1}(\epsilon_i) + \mathcal{O}\left(\frac{\log(n_i)}{n_i}\right),
    \label{eq:3}
\end{equation}
where $C(x) \triangleq \log (1 + x)$ is the channel capacity\footnote{$\ln(\cdot)$ and $\log(\cdot)$ denote the natural and base-$2$ logarithms, respectively.}, $V(x) \triangleq 2 x(1 + x)^{-1}\log^2 e$ is the so-called channel dispersion \cite{Sca17}, $Q(\cdot)$ is the Gaussian Q-function, $\epsilon_i$ is the error probability, and $n_i$ is the number of transmit symbols. In fact, for a sufficiently large $n_i$, the last addend in \eqref{eq:3} can be safely omitted.

Let $a(n_i,\epsilon_i) \triangleq (\log e / \sqrt{n_i}) Q^{-1}(\epsilon_i)$ and $\Delta(\rho_i) \triangleq \sqrt{V(\rho_i)}$ such that $R(\rho_i,n_i,\epsilon_i) \approx C(\rho_i) - a(n_i,\epsilon_i) \Delta(\rho_i)$. We consider that $n_i$ and $\epsilon_i$ are fixed and detailed by the system requirements (specified in Section~\ref{sec:4}) and, thus, the data rate only depends on the SINR or, equivalently, on $\bm{\psi}$, i.e., $R_i(\bm{\psi}) \equiv R(\rho_i,n_i,\epsilon_i)$, $C_i(\bm{\psi}) \equiv C(\rho_i)$, $a_i \equiv a(n_i,\epsilon_i)$, and $\Delta_i(\bm{\psi}) \equiv \Delta(\rho_i)$. 

Based on that, two interesting optimization problems can be formulated, namely maximization of the weighted sum rate (WSR) and maximization of the minimum rate:
\begin{gather}
\label{eq:4} \bm{\psi}^{\star} = \underset{\bm{\psi}}{\textrm{argmax}} \, \, \sum_{i=1}^M \omega_i R_i(\bm{\psi}) \quad \textrm{s.t.} \quad   \vert \psi_l \vert \leq 1 \, \forall l, \\
\label{eq:5} \bm{\psi}^{\star} = \underset{\bm{\psi}}{\textrm{argmax}} \, \underset{i}{\textrm{min}} \, \, R_i(\bm{\psi}) \quad \textrm{s.t.} \quad   \vert \psi_l \vert \leq 1 \, \forall l,
\end{gather}
where $\omega_i$ are the different priority weights \cite{Guo20}. Unfortunately, the objective functions are not convex in $\bm{\psi}$. That is why in the following, we propose methods to find a practical solution\footnote{The widely used hard-equality constraint $\vert \psi_l \vert = 1$ is a particular case and results in equal or worse performance (due to the smaller feasible set).}. 

\section{Proposed Solutions} \label{sec:3}
To tackle the previous optimizations, in this section, we will explore two different strategies: (i) GA and (ii) SO with SDR. Both approaches are based on an iterative (successive) procedure, where at each step (denoted by $k$) the solution $\bm{\psi}^{(k)}$ is updated (refined) until convergence is achieved. In (i), we distinguish between the Euclidean and Riemannian gradients. 

\subsection{Gradient Ascent} \label{sec:3.1}
Given that the throughput is continuous and differentiable, the first approach to finding a sub-optimal RIS configuration is the well-known GA algorithm \cite{Boy04}. Considering the WSR maximization\footnote{The minimum rate solution can be derived using similar techniques.}, the solution can be found as follows\cite{Che18}:
\begin{equation}
    \bm{\psi}^{(k + 1)} = \mathcal{P}\left(\bm{\psi}^{(k)} + \alpha_k \sum_{i = 1}^M \omega_i \textrm{grad} \, R_i\left(\bm{\psi}^{(k)}\right)\right),
    \label{eq:6}
\end{equation}
with $\mathcal{P}(\cdot)$ the operator projecting the solution into the subspace defined by the constraint set\footnote{The constraint set can be satisfied by, for example, normalizing the vector $\bm{\psi}$ by its largest absolute value, i.e., $\bm{\psi} \gets \bm{\psi}/a$ with $a \triangleq \textrm{max}_l \vert \psi_l \vert$.} and $\alpha_k > 0$ the Armijo step size \cite{Guo20}. To ease of notation, $\textrm{grad} \, R_i\left(\bm{\psi}^{(k)}\right)$ can refer either to the Euclidean or to the Riemannian gradient of the data rate. In short, the Riemannian case is an extension of the Euclidean as, despite the projection, it considers the geometric properties of the constraint set in the search \cite{Che18}. As shown below, both expressions can be derived using complex vector calculus \cite{Pet12}.

\subsubsection{Euclidean Gradient}
\begin{equation}
    \textrm{grad} \, R_i\left(\bm{\psi}^{(k)}\right) = \nabla R_i\left(\bm{\psi}^{(k)}\right) = \nabla C_i\left(\bm{\psi}^{(k)}\right) - a_i \nabla \Delta_i\left(\bm{\psi}^{(k)}\right),
    \label{eq:7}
\end{equation}
where $\nabla R_i\left(\bm{\psi}^{(k)}\right)$ denotes the Euclidean gradient of $R_i\left(\bm{\psi}^{(k)}\right)$,
\begin{equation}
    \nabla C_i\left(\bm{\psi}^{(k)}\right) = \frac{2}{\ln 2} \left(\frac{P_i \bm{h}_i\bm{h}_i^{\textrm{H}}\left(\bm{\psi}^{(k)}\right)^* - \Omega_{i}\left(\bm{\psi}^{(k)}\right)}{\sigma_w^2 + \sum_{j \geq i} P_j \vert \left(\bm{\psi}^{(k)}\right)^{\textrm{T}}\bm{h}_j \vert^2}\right) ,
    \label{eq:8}
\end{equation}
represents the Euclidean gradient of $C_i\left(\bm{\psi}^{(k)}\right)$ with
\begin{equation}
\begin{split}
    \Omega_i\left(\bm{\psi}^{(k)}\right) &\triangleq \frac{P_i \vert \left(\bm{\psi}^{(k)}\right)^{\textrm{T}}\bm{h}_i \vert^2\sum_{j > i} P_j \bm{h}_j\bm{h}_j^{\textrm{H}}\left(\bm{\psi}^{(k)}\right)^*}{\sigma_w^2 + \sum_{j > i} P_j \vert \left(\bm{\psi}^{(k)}\right)^{\textrm{T}}\bm{h}_j \vert^2},
    \label{eq:9}
    \end{split}
\end{equation}
and
\begin{equation}
    \nabla \Delta_i\left(\bm{\psi}^{(k)}\right) = \sqrt{2\frac{\sigma_w^2 + \sum_{j \geq i} P_j \vert \left(\bm{\psi}^{(k)}\right)^{\textrm{T}}\bm{h}_j \vert^2}{ P_i \vert \left(\bm{\psi}^{(k)}\right)^{\textrm{T}}\bm{h}_i \vert^2}} \Lambda_{i}\left(\bm{\Phi}^{(k)}\right),
    \label{eq:10}
\end{equation}
is the Euclidean gradient of $\Delta_i\left(\bm{\psi}^{(k)}\right)$ with
\begin{equation}    
    \Lambda_i\left(\bm{\psi}^{(k)}\right) \triangleq \frac{P_i \bm{h}_i\bm{h}_i^{\textrm{H}}\left(\bm{\psi}^{(k)}\right)^*}{\sigma_w^2 + \sum_{j \geq i} P_j \vert \left(\bm{\psi}^{(k)}\right)^{\textrm{T}}\bm{h}_j \vert^2} - \frac{P_i \vert \left(\bm{\psi}^{(k)}\right)^{\textrm{T}}\bm{h}_i \vert^2\sum_{j \geq i} P_j \bm{h}_j\bm{h}_j^{\textrm{H}}\left(\bm{\psi}^{(k)}\right)^*}{(\sigma_w^2 + \sum_{j \geq i} P_j \vert \left(\bm{\psi}^{(k)}\right)^{\textrm{T}}\bm{h}_j \vert^2)^2}.
    \label{eq:11}    
\end{equation}

\subsubsection{Riemannian Gradient}
\begin{equation}
    \textrm{grad} \, R_i\left(\bm{\psi}^{(k)}\right) = \nabla R_i\left(\bm{\psi}^{(k)}\right) \nonumber - \mathrm{Re}\left\{\nabla R_i\left(\bm{\psi}^{(k)}\right) \odot \left(\bm{\psi}^{(k)}\right)^*\right\} \odot \bm{\psi}^{(k)}, \label{eq:12}
\end{equation}
where $\odot$ is the Hadamard (or element-wise) product.

\subsection{Sequential Optimization with Semi-Definite Relaxation}
Before applying SDR, we start this subsection by finding a tight concave lower bound for the data rate $R_i\left(\bm{\psi}^{(k)}\right)$. Later on, this new objective function will be sequentially optimized (maximized) until a stationary solution is found \cite{Hua19}.

To do so, we will consider the following inequalities \cite{Nas21}:
\begin{gather}
    \label{eq:13} \ln \left(1 + \frac{x}{y}\right) \geq \ln \left(1 + \frac{\bar{x}}{\bar{y}}\right) + \frac{\bar{x}}{\bar{y}} \left(2\sqrt{\frac{x}{\bar{x}}} - \frac{x + y}{\bar{x} + \bar{y}} - 1\right), \\
    \label{eq:14} \sqrt{x} \leq \frac{1}{2} \left(\sqrt{\bar{x}} + \frac{x}{\sqrt{\bar{x}}}\right),
    \quad x \leq \frac{1}{2} \left(\bar{x} + \frac{x^2}{\bar{x}}\right),
\end{gather}
for $x > 0$, $y > 0$, $\bar{y} > 0$, and $\bar{x} > 0$.

We will first find a concave lower bound for $C_i\left(\bm{\psi}^{(k)}\right)$ and then a convex upper bound for $\Delta_i\left(\bm{\psi}^{(k)}\right)$. In that sense, the capacity bound at step $k$ follows from \eqref{eq:13}:
\begin{equation}
    C_i\left(\bm{\psi}^{(k)}\right) \geq \frac{1}{\log 2} \left(\ln \left(1 + \rho_i^{(k - 1)}\right) + \rho_i^{(k - 1)} \Gamma_i \left(\bm{\psi}^{(k)}\right)\right),
\label{eq:15}
\end{equation}
where
\begin{equation}    
    \Gamma_i \left(\bm{\psi}^{(k)}\right) \triangleq 2\sqrt{\frac{ \left(\bm{\psi}^{(k)}\right)^{\textrm{T}} \bm{H}_i \left(\bm{\psi}^{(k)}\right)^{*} }{\left(\bm{\psi}^{(k - 1)}\right)^{\textrm{T}} \bm{H}_i \left(\bm{\psi}^{(k - 1)}\right)^{*}}} - \frac{\sigma_w^2 + \left(\bm{\psi}^{(k)}\right)^{\textrm{T}} \bar{\bm{H}}_i\left(\bm{\psi}^{(k)}\right)^{*}}{\sigma_w^2 + \left(\bm{\psi}^{(k - 1)}\right)^{\textrm{T}} \bar{\bm{H}}_i\left(\bm{\psi}^{(k - 1)}\right)^{*}} - 1,
    \label{eq:16}    
\end{equation}
with $\bm{H}_i \triangleq P_i \bm{h}_i \bm{h}_i^{\textrm{H}}$, $\tilde{\bm{H}}_i \triangleq \sum_{j > i} \bm{H}_j$, and $\bar{\bm{H}}_i\triangleq \bm{H}_i + \tilde{\bm{H}}_i$. Note that $\rho_i^{(k - 1)}$ in \eqref{eq:15} is the SINR from the previous iteration:
\begin{equation}
    \rho_i^{(k - 1)} =  \frac{\left(\bm{\psi}^{(k - 1)}\right)^{\textrm{T}} \bm{H}_i \left(\bm{\psi}^{(k - 1)}\right)^{*} }{\sigma_w^2 + \left(\bm{\psi}^{(k - 1)}\right)^{\textrm{T}} \tilde{\bm{H}}_i \left(\bm{\psi}^{(k - 1)}\right)^{*}}.
    \label{eq:17}
\end{equation}

Unfortunately, since $\bm{H}_i$ are positive (semi-) definite matrices, the set of functions $\sqrt{\left(\bm{\psi}^{(k)}\right)^{\textrm{T}} \bm{H}_i \left(\bm{\psi}^{(k)}\right)^{*}}$ are not concave in $\bm{\psi}^{(k)}$. Hence, although $\left(\bm{\psi}^{(k)}\right)^{\textrm{T}} \bar{\bm{H}}_i\left(\bm{\psi}^{(k)}\right)^{*}$ is indeed convex, the previous bound results non-concave.

A possible alternative is to define $\bm{\Phi}^{(k)} \triangleq \left(\bm{\psi}^{(k)}\right)^{*} \left(\bm{\psi}^{(k)}\right)^{\textrm{T}}$ and reformulate the optimization in terms of $\bm{\Phi}^{(k)}$ \cite{Wu19}. As a result, the set of (convex) constraints $  \vert \psi_l \vert \leq 1 \, \forall l$ translate into
\begin{equation}
\mathcal{C} \triangleq \left\{\bm{\Phi}^{(k)} \succeq \bm{0}, \textrm{rank}\left(\bm{\Phi}^{(k)}\right) = 1,   \left[\bm{\Phi}^{(k)}\right]_{l,l} \leq 1 \, \forall l\right\}. 
\label{eq:18}
\end{equation}

Accordingly, the bound in \eqref{eq:15} becomes
\begin{align}
    C_i\left(\bm{\Phi}^{(k)}\right) &= \log \left(1 + \frac{\textrm{tr}\left(\bm{\Phi}^{(k)}\bm{H}_i\right)}{\sigma_w^2 + \textrm{tr}\left(\bm{\Phi}^{(k)}\tilde{\bm{H}}_i\right)}\right) \nonumber \\ 
    &\geq C_i\left(\bm{\Phi}^{(k - 1)}\right) + \frac{1}{\log 2} \frac{\textrm{tr}\left(\bm{\Phi}^{(k - 1)}\bm{H}_i\right) \Gamma_i\left(\bm{\Phi}^{(k)}\right)}{\sigma_w^2 + \textrm{tr}\left(\bm{\Phi}^{(k - 1)}\tilde{\bm{H}}_i\right)}  \nonumber \\
    &\triangleq \tilde{C_i}\left(\bm{\Phi}^{(k)}\right),
\label{eq:19}
\end{align}
which can be shown to be concave with respect to (w.r.t.) $\bm{\Phi}^{(k)}$. Indeed, the term $\Gamma_i\left(\bm{\Phi}^{(k)}\right)$ is given below:
\begin{equation}
    \Gamma_i\left(\bm{\Phi}^{(k)}\right) \triangleq 2\sqrt{\frac{ \textrm{tr}\left(\bm{\Phi}^{(k)}\bm{H}_i\right) }{\textrm{tr}\left(\bm{\Phi}^{(k - 1)}\bm{H}_i\right)}} - \frac{\sigma_w^2 + \textrm{tr}\left(\bm{\Phi}^{(k)}\bar{\bm{H}}_i\right)}{\sigma_w^2 + \textrm{tr}\left(\bm{\Phi}^{(k - 1)}\bar{\bm{H}}_i\right)} - 1,
    \label{eq:20}
\end{equation}
which is the sum of a concave function (square root of a linear function) and a linear function; therefore, it is concave.

Regarding the term $\Delta_i\left(\bm{\Phi}^{(k)}\right)$, we can obtain a convex upper bound by means of the inequalities in \eqref{eq:14}:
\begin{align}
    \Delta_i\left(\bm{\Phi}^{(k)}\right) &= \sqrt{\frac{2 \textrm{tr}\left(\bm{\Phi}^{(k)} \bm{H}_i\right) }{\sigma_w^2 + \textrm{tr}\left(\bm{\Phi}^{(k)}\bar{\bm{H}}_i\right)}} \nonumber \\ 
    &\leq \frac{\Delta_i\left(\bm{\Phi}^{(k - 1)}\right)}{2} + \frac{1}{\Delta_i\left(\bm{\Phi}^{(k - 1)}\right)}\frac{ \textrm{tr}\left(\bm{\Phi}^{(k)}\bm{H}_i \right) }{\sigma_w^2 + \textrm{tr}\left(\bm{\Phi}^{(k)}\bar{\bm{H}}_i\right)} \nonumber \\
    &\leq \frac{\Delta_i\left(\bm{\Phi}^{(k - 1)}\right)}{2} + \frac{ \textrm{tr}\left(\bm{\Phi}^{(k -1)}\bm{H}_i \right)}{2 \Delta_i\left(\bm{\Phi}^{(k - 1)}\right)} \frac{1}{ \sigma_w^2 + \textrm{tr}\left(\bm{\Phi}^{(k)}\bar{\bm{H}}_i\right)} \nonumber \\ 
    &\quad + \frac{1}{2 \Delta_i\left(\bm{\Phi}^{(k - 1)}\right)\textrm{tr}\left(\bm{\Phi}^{(k -1)}\bm{H}_i \right)}\frac{\textrm{tr}^2\left(\bm{\Phi}^{(k )}\bm{H}_i \right)}{\sigma_w^2 + \textrm{tr}\left(\bm{\Phi}^{(k)}\bar{\bm{H}}_i\right)} \nonumber \\
    &\triangleq \tilde{\Delta}_i\left(\bm{\Phi}^{(k)}\right),
\label{eq:21}
\end{align}
where the last addend is quadratic over linear, thus convex.

Finally, the new optimization problems read as
\begin{gather}
\label{eq:22} \bm{\Phi}^{(k)} = \underset{\bm{\psi}}{\textrm{argmax}} \, \, \sum_{i=1}^M \omega_i \tilde{R}_i\left(\bm{\Phi}^{(k)}\right) \quad \textrm{s.t.} \quad \mathcal{C}, \\
\label{eq:23} \bm{\Phi}^{(k)} = \underset{\bm{\psi}}{\textrm{argmax}} \, \underset{i}{\textrm{min}} \, \, \tilde{R}_i\left(\bm{\Phi}^{(k)}\right) \quad \textrm{s.t.} \quad \mathcal{C},
\end{gather}
where $\tilde{R}_i\left(\bm{\Phi}^{(k)}\right)$ is the concave lower bound of the rate in \eqref{eq:3}:
\begin{equation}
    \tilde{R}_i\left(\bm{\Phi}^{(k)}\right) = \tilde{C_i}\left(\bm{\Phi}^{(k)}\right) - a_i \tilde{\Delta}_i\left(\bm{\Phi}^{(k)}\right).
    \label{eq:24}
\end{equation}

By applying SDR and dropping the (non-convex) rank-one constraint, the above problems can be solved using standard numerical methods, e.g., CVX \cite{CVX20}. Finally, once convergence is reached, the rank-one solution can be retrieved with the help of Gaussian randomization (although a stationary point cannot be then guaranteed) \cite{Wu19}. This is discussed in the simulations.

\subsection{Complexity Analysis}
The computational cost (per iteration) of the GA-based approaches is mainly given by the computation of the Euclidean gradient, i.e., $\mathcal{O}(M^2L^2)$ \cite{Guo20}. Contrarily, due to the convex formulation, the complexity of the SO technique is polynomial w.r.t. $L^2$ (the number of variables) \cite{Boy04}. The analysis of the iterations needed by each approach is reserved for Section~\ref{sec:4}.

\section{Numerical Simulations} \label{sec:4}
To assess the performance of the proposed approaches, here we will present the resulting data rates w.r.t. the number of reflecting elements $L$. For a broader comparison, the situation where the RIS elements $\psi_l$ are found sequentially using one-dimensional exhaustive searches (AO) is also included \cite{Wu19}.

Throughout this section, we will consider the micro-urban scenario described in \cite{ITU09} with $P_i = 0$ dBm $\forall i$, $\sigma_w^2 = N_o B$, $N_o = -174$ dBm/Hz, and $B = 1.08$ MHz \cite{3GPP45820}. We assume a setup with $M = 10$ sensors uniformly distributed around the RIS within a disk of radius $10$ m. All channels are modeled using a power-law path loss and a Rician fading with factors $10$ and $1$ for $\bm{g}_{\textrm{R}}$ and $\bm{g}_i$, respectively. The steering vectors are generated for a uniform planar array configuration \cite{Wan21}. We also consider $n_i = 100$ symbols $\forall i$ and $\epsilon_i = 10^{-3}$ $\forall i$.

\begin{figure}[t]
    \centering
    \includegraphics[scale = 1.5]{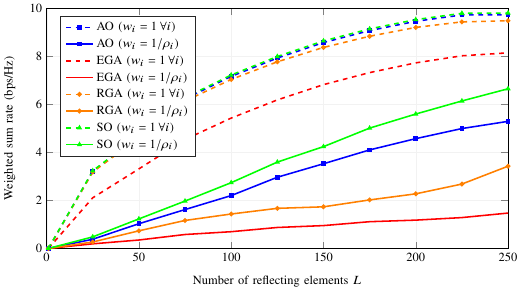}
    \caption{WSR obtained with the solution from \eqref{eq:22} versus $L$.}    
    \label{fig:2}
\end{figure}

The WSR is depicted in Fig.~\ref{fig:2} for different weight criteria: equality ($\omega_i = 1$ $\forall i$) and fairness ($\omega_i = 1/\rho_i$). All weights are normalized so that $\sum_i \omega_i = M$. The SO method provides the highest performance in the fair setup, but similar results are obtained with the AO, Riemannian GA (RGA), and SO when choosing equal priorities. This is because the solution in the case of equal weights can be found easily (e.g., capitalizing on the devices with the best conditions). Contrarily, the Euclidean GA (EGA) always yields the poorest values.

\begin{figure}[t]
    \centering    
    \includegraphics[scale = 1.5]{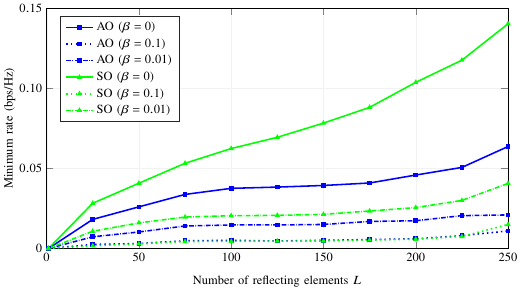}
    \caption{Minimum rate obtained with the solution from \eqref{eq:23} versus $L$.}
    \label{fig:3}
\end{figure}

The minimum rate is illustrated in Fig.~\ref{fig:3} for the AO and SO solutions only (the gradient-based searches are omitted to avoid redundancy). As before, the SO strategy outperforms the AO technique, especially for large $L$. Additionally, to study the impact of imperfect channel knowledge, we include the case where the actual value $\bm{h}_i$ is contaminated by an estimation error $\bm{\eta}_i \sim \mathcal{CN}(\mathbf{0}_L, \sigma_{\eta}^2\mathbf{I}_L)$, where $\sigma_{\eta}^2 = \frac{\beta}{L} \textrm{tr}(\bm{C}_i)$, $\beta \in [0,1]$, and $\bm{C}_i \in \mathbb{C}^{L \times L}$ is the covariance matrix of $\bm{h}_i$. As expected, the performance degrades significantly with $\sigma_{\eta}^2$ (or $\beta$).

\begin{figure}[t]
    \centering    
    \includegraphics[scale = 1.5]{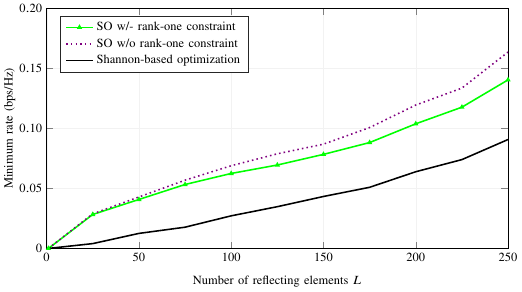}
    \caption{Minimum rates obtained with the solution from \eqref{eq:23} and from the maximization of the minimum Shannon capacity versus $L$.}
    \label{fig:4}
\end{figure}

In Fig.~\ref{fig:4}, we illustrate the previous minimum rate $R_i$ and that obtained when maximizing the minimum Shannon capacity $C_i$ through SO (the WSR is omitted to avoid redundancy). For the latter, the RIS is designed by considering $n_i \to \infty$ in \eqref{eq:23}. Since the capacity is independent of the blocklength and does not include the dispersion term (cf. \eqref{eq:3}), there will be a mismatch between the metrics $R_i$ and $C_i$. As we can see, this results in a large performance gap that can be even wider with shorter packets (i.e., $n_i < 100$). Hence, the optimization of the Shannon bound can be misleading for mMTC transmissions and the FBLR analysis becomes imperative.

To further benchmark the performance of the SO, we also include the throughput with (w/-) and without (w/o) the rank-one constraint. Since both curves can be quite tight, one can state that in this case, the SO with SDR is (nearly) optimal. 

Note that the previous reasoning also holds for the case of gathering RIS elements. Considering groups of size $G \geq 1$, $L$ would refer to the number of groups and the actual number of unit cells would be $L G \geq L$ \cite{Dir22}. However, the analysis inclu\-ding grouping techniques is beyond the scope of this work. 

\begin{figure}[t]
    \centering    
    \includegraphics[scale = 1.5]{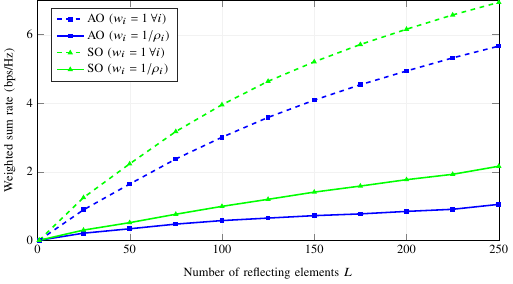}
    \caption{WSR obtained with the solution from \eqref{eq:23} versus $L$.}
    \label{fig:5}
\end{figure}

\begin{figure}[t]
    \centering    
    \includegraphics[scale = 1.5]{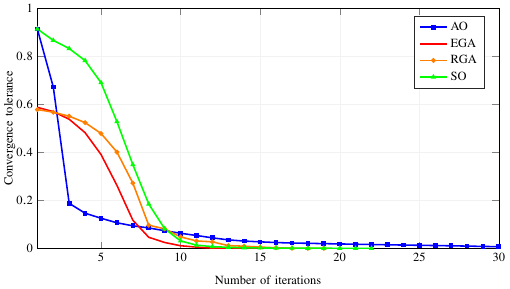}
    \caption{Convergence tolerance versus the number of iterations.}
    \label{fig:6}
\end{figure}

\begin{table}[t]
\caption{Execution time (in seconds).}
\begin{center}
\begin{tabular}{|c|c|c|c|c|c|c|c|}
\hline
AO & EGA & RGA & SO \\ 
\hline
$618.84$ & $0.083$ & $0.18$ & $327.61$ \\
\hline
\end{tabular}
\label{tab:1}
\end{center}
\end{table}

To get richer insights, the WSR obtained with the minimum rate solution is depicted in Fig.~\ref{fig:5}. This illustrates the trade-off between both criteria, i.e., maximizing the minimum throughput deteriorates the WSR and vice versa. In addition to that, compared to Fig.~\ref{fig:2}, now the results with equal weights follow a similar behavior to those with distinct priorities: the SO greatly outperforms the AO. The reason behind this fact is that, apart from finding all variables jointly (not separately as in AO), our proposal stands out in fair optimizations.

Finally, regarding the complexity of the proposed methods, in Fig.~\ref{fig:6} and Table \ref{tab:1} we present the convergence tolerance (i.e., the relative difference between current and final rates) versus the number of iterations $k$ and the corresponding execution time for the fair WSR optimization with $L = 100$ reflecting elements, respectively. From all approaches, the EGA is the one with the smallest number of iterations whereas the AO requires the highest execution time for convergence.

\section{Conclusions} \label{sec:5}
In this paper, we have addressed the problem of designing a RIS for the maximization of the FBLR data rate in an mMTC network. In a setup where sensors transmit their information to a CN, we have presented two feasible strategies to find a sub-optimal configuration of the RIS. The first is based on the popular GA, for which Euclidean and Riemannian gradients are considered. The other approach relies on the derivation of a concave lower bound for the data rate and the use of SO with SDR. Simulations highlight the performance of the SO and the importance of the FBLR analysis in mMTC transmissions.

\bibliographystyle{IEEEtran}
\bibliography{IEEEabrv,References}
\end{document}